\documentclass[conference]{IEEEtran}
\makeatletter
\def\ps@headings{%
\def\@oddhead{\mbox{}\scriptsize\rightmark \hfil \thepage}%
\def\@evenhead{\scriptsize\thepage \hfil \leftmark\mbox{}}%
\def\@oddfoot{}%
\def\@evenfoot{}}
\makeatother
\IEEEoverridecommandlockouts
\usepackage{cite}
\usepackage{graphicx}
\usepackage[cmex10]{amsmath}
\usepackage{amssymb}
\usepackage{url}
\usepackage{subfigure}
\usepackage{algorithmic}
\usepackage{algorithm}
\usepackage{array}

\begin{document}

\title{On the Benefit of Information Centric Networks for Traffic Engineering}

\author{\IEEEauthorblockN{Kai Su}
\IEEEauthorblockA{ WINLAB\\
Rutgers University\\
North Brunswick, NJ\\
kais@winlab.rutgers.edu }
\and
\IEEEauthorblockN{Cedric Westphal\hspace{0.5cm}}
\IEEEauthorblockA{ \hspace{1cm}Innovation Center \hspace{2.5cm}Dept. of Computer Engineering\hspace{0.5cm}\\ Huawei Technology \hspace{3cm}University of California\\Santa Clara, CA\hspace{4cm}Santa Cruz, CA\\
\hspace{2cm}cedric.westphal@huawei.com \hspace{2.5cm}cedric@soe.ucsc.edu\hspace{2.5cm}}
}

%
\maketitle

\begin{abstract}
Current Internet performs traffic engineering (TE) by estimating traffic matrices on a regular schedule, and allocating flows based upon weights computed from these matrices. This means the allocation is based upon a guess of the traffic in the network based on its history. Information-Centric Networks on the other hand provide a finer-grained description of the traffic: a content between a client and a server is uniquely identified by its name, and the network can therefore learn the size of different content items, and perform traffic engineering and resource allocation accordingly. We claim that Information-Centric Networks can therefore provide a better handle to perform traffic engineering, resulting in significant performance gain.

We present a mechanism to perform such resource allocation. We see that our traffic engineering method only requires knowledge of the flow size (which, in ICN, can be learned from previous data transfers) and outperforms a min-MLU allocation in terms of response time. We also see that our method identifies the traffic allocation patterns similar to that of min-MLU without having access to the traffic matrix ahead of time. We show a very significant gain in response time where min MLU is almost 50\% slower than our ICN-based TE method. 
\end{abstract}


\section{Introduction}

Finer grained resource allocation is primordial to achieve better link utilization and network performance~\cite{Jain2013B4}\cite{Hong2013Achieving}. However, current IP networks suffer from a lack of predictability of the traffic patterns which makes resource allocation difficult. Traffic flow lengths in the Internet follow a power law distribution (see for instance~\cite{Clegg2010Critical} and references therein). Assigning flows to links becomes suboptimal if multiple long-lived flows end up sharing the same bottleneck resource. 

The typical link allocation is based upon the recent past of the network traffic: the network monitors the amount of traffic during a given time period, measures the flow in between ingress and egress points, and computes a traffic matrix for this time period. This matrix is then used to decide how to allocate flows to resource (typically by associating set of IP destination addresses to MPLS labels) over the next time period. The typical goal is to minimize the maximum link utilization (MLU) so as to split the traffic as evenly as possible across the network, with the hope that traffic patterns will resemble that of the previous period.

Information-centric networks (ICN,~\cite{Xylomenos2013Survey}\cite{Ahlgren2012Survey}\cite{Fayazbakhsh2013LessPain}) have been proposed as a way to assist the Internet in coping with the explosion of bandwidth demand by facilitating content distribution (which is predicted to amount to 90\% of the Internet traffic by 2017 by the Cisco® Visual Networking Index). One key ingredient of ICN architectures is that the content is uniquely mapped to a globally identifiable name. This in turn allows the network to discover properties of the content and to associate these properties as attributes to the name. Therefore, it is possible to know ahead of time if a specific piece of content will be a mouse or an elephant. This shifts the allocation problem from open-ended semantics, where one can only try to guess how much resource a specific flow will consume, to deterministic semantics. 

\cite{chanda2013content}  described a mechanism for the network to learn the size of content going through the networks and suggested that a network controller could assign resources to content based upon the content size. In this paper, we describe such a resource allocation mechanism. We show how a dynamic traffic engineering could select paths based upon specific content attributes, and compare this with a typical min-MLU traffic engineering policy. 

Most literature describing the potential benefits of ICN focus on caching and content delivery. We argue here that ICNs actually perform significantly better in terms of traffic engineering when compared with a content-oblivious min MLU traffic engineering policy. Indeed, min MLU's response time in our evaluation is 30\% to 50\% longer than our content-based traffic engineering. The policy we propose is relatively simple, and we hope that better policies will be proposed in the future. However, it does make the point that content-awareness has a strong impact on network performance. 

The paper is organized as follows. In section~\ref{sec:rel-work}, we present briefly some prior literature on the topic. We discuss our design considerations in Section~\ref{sec:design-space} and describe our traffic engineering policy in Section~\ref{sec:MBP}. We thoroughly evaluate the policy in Section~\ref{sec:eval} and provide some concluding remarks in Section~\ref{sec:conclusion}.

\section{Related work} \label{sec:rel-work}

Information-centric networks~\cite{Xylomenos2013Survey}\cite{Ahlgren2012Survey}\cite{Fayazbakhsh2013LessPain} currently focus on how to securely name data, how to route based upon names, and where to cache content in the network. The issue of traffic engineering has come up in relatively few ICN-related works. For instance, TECC~\cite{Xie2012TECC} considers how to jointly provision content routing and caching capabilities. However, it solves some optimization problem based upon traffic matrices to assign costs to link off-line, not by taking into account the network conditions as we do.

\cite{chanda2013content} identified the potential for finer grained resource allocations using an ICN architecture, but it stayed short of providing a traffic engineering policy and conducted only a cursory evaluation upon a two-link topology.

For traffic engineering in IP networks, \cite{Fortz2000Internet} computed OSPF link weights to allocate IP traffic. Current architectures have no way to identify the content directly at the network layer, and rely on CDN heuristics\cite{Yu2012Tradeoffs} to improve the performance of content distribution. \cite{Sharma2013Distributing} discusses the benefit of performing traffic engineering jointly with the content distribution strategy in a ``network-CDN'', but considers only  history-based strategy. 

\cite{Poese2012Enabling} demonstrates the benefit of CDN server selection performed in conjunction with the ISP's recommendation over short time scales. Our set-up is different in the sense that we assume the content location is chosen independently of the network operator. Namely, the copy of the content can be hosted outside of the network's domain and only transits through the network, or the CDN is not cooperating with the ISP. The philosophy is similar to that of~\cite{Jiang2009Cooperative} or~\cite{DiPalantino2009Traffic} which both consider joint traffic engineering and content selection using game theoretic frameworks.

\cite{Sharma2011Beyond} demonstrates the need for better TE tools by comparing MLU with other TE schemes as they impact application performance. \cite{Wang2006COPE} also attempts to replace min-MLU schemes in order to take into account the unpredictability of the traffic. In our scheme, predictability is improved by obtaining content metadata at the network layer, allowing a finer grained resource allocation. 

Our policy is based on a minimal backlog allocation, which has been used in previous context of parallel servers and analytically described in\cite{Dai2007Stability} or \cite{Gupta2007Analysis}. There is no such theory for networks of queues beyond the case of parallel queues, unfortunately.

\section{Design space for the TE algorithm}
\label{sec:design-space}

\subsection{Network assumptions}

We consider the following set-up: flows associated with a content arrive into the network (either generated within the domain, or arriving at a peering point from a neighboring domain). Each flow is a series of packets or chunks that correspond to a specific piece of data. Upon the arrival of such a flow, the network decides how to allocate the flow to a path in the network. 

As in Information-Centric Networks, each packet carries a name that uniquely identifies the content that the flow carries, in addition to some extra information to disambiguate the multiple chunks or segments which compose the content. 

We consider that the node making the allocation has a global view of the domain, where the view is a measure of the congestion in the network which we will define later. It could be that the ingress node into the domain is aware of this congestion measure for the different paths to the different possible destinations, or that it invokes a logically centralized controller, as in software-defined networks. For clarity of exposition, we assume from now on that a controller makes the decision, but the process could be decentralized to the ingress nodes in the network.

Upon the arrival of a flow, the network performs the allocation based upon its awareness of the congestion level over the paths that the flow can take in the network. We assume that the decision is content-based (or flow-based, as flows are associated to content), namely that all the subsequent chunks of a flow will follow the same path. 

We also assume that the network extracts flow information (i.e., content size), and maintains a database with this information, as in~\cite{chanda2013content}. Alternatively, since Information-Centric Network architectures are not fully specified yet,  the flow could carry in the header this information as a new field. \cite{chanda2013content} describes a possible instantiation of the assumptions we are making, where a centralized control gather meta-data about the content into a database, and looks up this database for allocating resource to a new flow.

Denote by $G=(V,E)$ the network graph, where $V$ is the set of nodes  and $E$ is the set of links. Each link $e$ has capacity $c_e$. For each flow $z_{s,d}$ entering the network at vertex $s$ and leaving at vertex $d$, we have a set of $K_{s,d}$ distinct paths $(P_{s,d}^k, k=1,\ldots,K_{s,d})$ to choose from, where a path is an acyclic sequence of links in $E$ going from $u$ to $v$. If link $e$ belongs to path $P_{s,d}^k$ for some $s,d,k$, we say that $e \in P_{s,d}^k$.  We assume $K_{s,d}$ is relatively low to simplify the allocation decision and the management complexity. In the evaluation, we consider $K_{s,d} = 3$. 

Flows from $s$ to $d$ are generated according to a Poisson process with rate $\lambda_{s,d}$. Since each flow corresponds to a piece of content, and since the networks has access to the content size, we can assume that, upon arrival of a flow $f$ in the network, the network has access to the flow size (which we also denote by $z$). We assume that the size $z$ is drawn from a known distribution (that of the distribution of the content size, which can be empirically measured) with mean $\bar{z}$. 

We further assume that the amount of traffic under the arrival rate $\lambda_{s,d}$ and the distribution for $z$ is stable and can be allocated to the paths $P_{s,d}^k$ in a manner such that the load allocated to each link is less (on average) than this link's capacity. Namely, we assume that there exist coefficients $\pi_{s,d}^k, 1,\ldots,K_{s,d}$ with $0 \leq \pi_{s,d}^k \leq 1$ and $\sum_k \pi_{s,d}^k = 1$, such that the link utilization $u_e$ of link $e$ satisfies the following {\em feasibility condition}:
\begin{eqnarray}
\forall e \in E, \ u_e  = \sum_{P_{s,d}^k: e \in P_{s,d}^k} \lambda_{s,d} \pi_{s,d}^k \bar{z} < c_e
\end{eqnarray}

Note that the matrix $\{\bar{z} \cdot  \lambda_{s,d}\}_{(s,d) \in V\times V}$ corresponds to the traffic matrix in the network, and the $\pi_{s,d}^k$ corresponds to a static traffic engineering decision. For instance, a possible traffic engineering policy could be to randomly split the flows arriving from $s$ to $d$ with probability $\pi_{s,d}^k$ onto the $K_{s,d}$ possible paths $P_{s,d}^k$. We denote by Weighted Random this random splitting policy where the choice of coefficient $\pi_{s,d}^k$ minimizes $\max_{e \in E} u_e$. This is the typical {\em min-MLU} traffic engineering policy.

Because we assume the network is stable, the {\em link utilization cannot be improved}. As all the traffic coming into the network leaves the network, the same total amount of traffic traverses the network under the Weighted Random policy or our proposed policy. Our focus therefore is on improving some other network performance metrics, namely how much time a flow will spend in the network. 

We define the {\em response time} of the network for flow $z$ as the time between the first arrival of $z$ at the ingress of the network until the last departure from the flow $z$ at the egress of the network.

Our focus is to reduce the response time of the network using the information made available to the network by using an Information-Centric architecture.

\subsection{Congestion Awareness}

We assume that each node on the path\footnote{A subset of the nodes might be sufficient to monitor traffic usage across the different paths. It is beyond the scope of this paper to identify a minimal subset to achieve this goal.} keeps track of the amount of traffic allocated to the node's outgoing links. Namely, when a flow $z_{s,d}$ with size $z$ is allocated to the $k$-th path $P_{s,d}^k$,  it will add $z$ to the backlog of the edges traversed by path $P_{s,d}^k$. 

The link $e$ is outgoing for node $w$, and $w$ keeps a backlog counter $B_e$ that is incremented every time a new flow is allocated to $e$ and which decreases by $c_e$ units per units of time (or how many bits are transmitted per unit of time, if it is less than $c_e$). $w$ reports $B_e$ to the nodes doing the resource allocation.

We assume that a transport mechanisms (say, TCP, or some form of interest-shaping in ICN) shares the bandwidth in a fair manner among the different flows using $e$. So the response time will not be affected by head-of-the-line blocking, but by how many flows are contending over a bottleneck resource.

\section{Minimum backlog policy design}
\label{sec:MBP}

In the current Internet, it is common that multiple distinct paths exist for a server and client pair. When a content of enormous size is to be delivered, it is anticipated that the corresponding flow would pose an enduring bandwidth demand on the links that it traverses. Thus care needs to be taken in the path selection process to prevent the incoming flow from congesting the heavily loaded links. In this section, we introduce the Minimum Backlog Policy (MBP) design which achieves this goal, taking into account the Information-Centric nature of the traffic.

\subsection{Motivating example} \label{subsec:moti}
\begin{figure}[t]
\label{fig:motiv}
\begin{center}

	\subfigure[Small network topology] {
		\label{fig:motiv-topo}
		 \includegraphics[height=1.5cm]{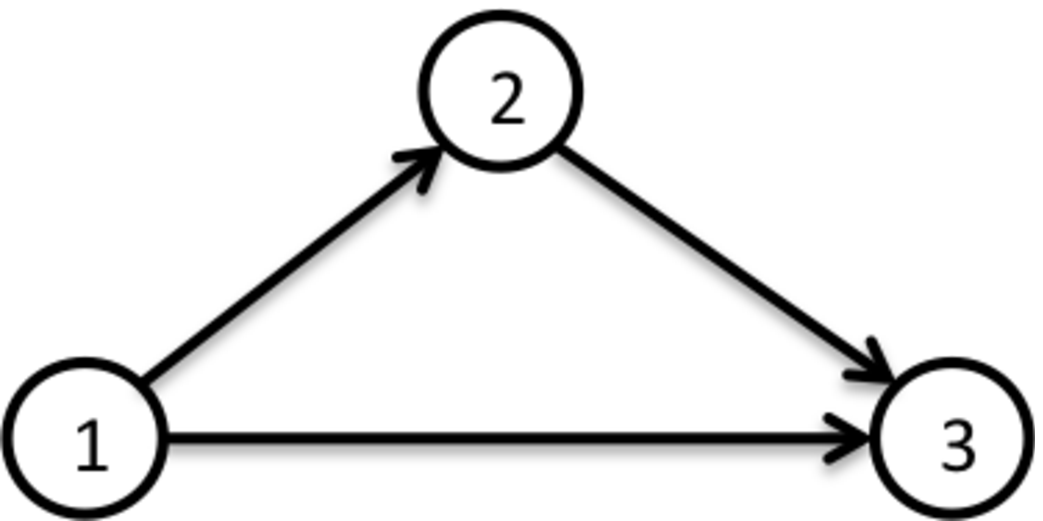}
	}
	\subfigure[Mean response time] {
		\label{fig:motiv-res}
		 \includegraphics[width=5cm]{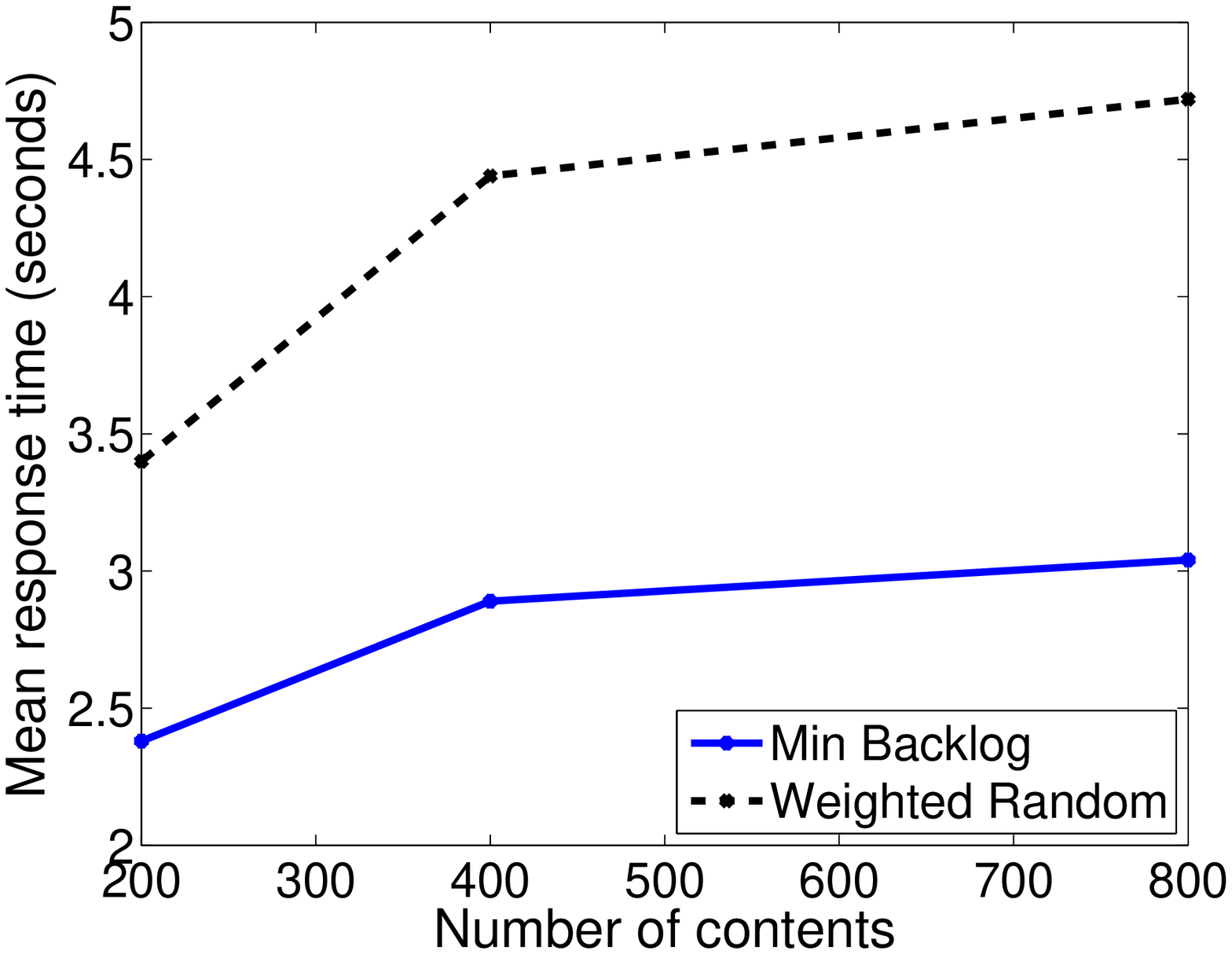}
	}
\caption{Motivating example: perform path selection for contents transferred from node 1 to node 3, based on two different policies, i.e., Min Backlog and Weighted Random, respectively.}
\end{center}
\end{figure}

We first consider a simple topology with $3$ nodes, shown in Figure \ref{fig:motiv-topo}, to motivate our design. Suppose each of the three links in this network has a capacity of $1$, and there exist $2$ traffic demands: (i) traffic of intensity $0.5$ from node $2$ to node $3$; (ii) traffic of intensity $0.5$ from node $1$ to node $3$. For the latter, two paths are available: the one traversing node $2$ and the direct one. We simulate this scenario by generating contents at node $1$ and node $2$ such that the specified traffic demands are satisfied. Suppose every time there is a content arrival at node $1$, the network decides which path it uses to deliver the content. We then perform an NS-2 event driven simulation for this simple case, regarding two different path selection policies: 

\begin{itemize}
\item \textbf{Weighted Random.}
Considering bottleneck link for the two paths have remaining capacity of $0.5$ and $1$, respectively, assign the flow to the first path with probability $1/3$, and to the other path with probability $2/3$. We call this policy Weighted Random. This is the min-MLU policy, as the link utilization is 0.75 on both link (2,3) and link (1,3). 
\item \textbf{Minimum Backlog.}
Assign the flow to the path whose maximum link backlog is less at the time of flow arrival. We call this policy Minimum Backlog Policy. This policy is unaware of the expected traffic demands, but requires knowledge of the network state.
\end{itemize}

Figure \ref{fig:motiv-res} shows the result when we simulate the network to send $800$ contents from node $1$ to node $3$. We measure the mean response time when $200, 400$ and $800$ contents are delivered, respectively. Figure \ref{fig:motiv-res} shows that MBP reduces the mean delay by a factor of about $30\%$ compared to Weighted Random (or conversely, that min-MLU is 50\% slower). 
\subsection{Minimum backlog policy design}
We now proceed to formulate the Minimum Backlog Policy for a general network topology. We consider a Content Centric Network as a directed graph, $G = (V,E)$. Suppose in this network, contents are being transferred from a set of content servers, $S \subset V$, to a set of requesting clients, $D \subset V$. We use $z_i$ to denote the $i$th content to be transferred in the network, as well as its size in bytes. For each point-to-point content transfer from the source $s \in S$ to the destination $d \in D$, we associate a backlog function with it. We use $B_{z_i}(t)$ to denote the backlog generated by the content $z_i$ from $s$ to $d$ at time $t$. Letting $t_{z_i}$ denote the arrival time of content $z_i$, then $B_{z_i}(t)$ is a non-increasing function of $t$ for $t\in [t_{z_i}, +\infty)$: it diminishes from the size of the content, $z_i$, down to $0$. For instance, if we consider a flow using the full capacity $c$ of a link, then $B_{z_i}(t)$ can be given as following:
\begin{equation}
B_{z_i}(t) = [z_i - c(t-t_{z_i})]^+.
\end{equation} 
where $[g]^+=\text{max}\{g, 0\}$. In general, due to the dynamics of the flow interactions and of the transport protocol (say, TCP), it is easier to compute $B_{z_i}(t)$ at each link by subtracting the volume of content that has traversed the link from the original content size. $B_{z_i}(t)$ correspond to the unfinished, remaining amount for the flow $z_i$.

The concept of backlog here is based on an end-to-end flow, and the actual backlog will only be held at the TCP sender's buffer. However, it should be noted that the remaining byte stream will eventually traverse all the links within a TCP connection. Since TCP throughput is limited by the bottleneck link, the larger the normalized size of the all the remaining byte streams combined at the bottleneck link, the longer the delay will be for the flow to finish if it traverses this link. Therefore we define the aggregated backlog at each link $e = (u,v)$,  $B_{(u,v)}$, at the time of $z_n$'s arrival, as the sum of remaining content to be transferred from all the tcp connections that traverse this link, normalized by the link capacity, i.e., 
\begin{equation}
B_{(u,v)} = \sum_{i=1}^{n-1} B_{z_i}(t) \cdot I(z_i,e) / c_{(u,v)}
\end{equation}
where $I(z_i,e)$ is an indicator function such that it equals $1$ if the transfer of content $z_i$ traverses link $e$, and $0$ otherwise, and $c_{u,v}$ is the capacity of link $(u,v)$. Thus when content $z_n$ arrives to be delivered, among a subset of all the paths between its source $s$ and destination $d$, $\mathcal P_{s,d}$, we attempt to find a path which achieves the following:
\begin{equation}
\begin{aligned}
& \underset{P \in \mathcal{P}_{s,d}}{\text{minimize}}
& & \underset{(p,q)\in P}{\text{max }} B_{(p,q)} \\
& \text{subject to}
&& B_{(u,v)} = \sum_{i=1}^{n-1} B_{z_i}(t) \cdot I(z_i,e),   \forall (u,v) \in E.
\end{aligned}
\end{equation}
i.e., to find a path whose maximum normalized link backlog is the least among $\mathcal P_{s,d}$. To keep this policy tractable, subsequently in our simulations, we specifically consider $\mathcal P_{s,d} = \{P_{s,d}^k, k=1,\ldots,K_{s,d}\}$ as the set of $K_{s,d}$ shortest paths given by the output of Yen's k-shortest path algorithm \cite{yen}.

Our algorithm for path selection for incoming content $z_n$, originated at node $a$ and destined for node $b$, is summarized in Algorithm \ref{alg:mbp}.

\begin{algorithm}
\caption{Minimum Backlog Policy}
\begin{algorithmic}
\REQUIRE $\mathcal P_{s,d}$ for each $(s, d)$ traffic demand pair. 
\STATE $B_{(i,j)} \leftarrow$ 0, $\forall (i,j) \in E$ 
\FOR{Every content $z_i$ being transferred, $i=1,\ldots,n-1$ \AND $B_{z_i}(t)>0$}
	\FOR{Every link $(u,v)$ in path $P$ carrying $z_i$} 
		\STATE $B_{(u,v)}(t) \ +=\  B_{z_i}(t)$
	\ENDFOR
\ENDFOR
\FOR{$P_{a,b}^k$ in $\mathcal P_{a,b}, k=1,\ldots,K_{a,b}$}
	\STATE $B_{P_{a,b}^k} = \underset{(u,v)\in P_{a,b}^k}{\text{max}} B_{(u,v)}$
\ENDFOR
\STATE $k^* = \underset{k}{\arg\min} \ B_{P_{a,b}^k}$
\RETURN path $P_{k^*}$
\end{algorithmic}
\label{alg:mbp}
\end{algorithm}

It might not be practical to compute the backlog for every content allocation. First, the estimation of the backlog $B_{(u,v)}(t)$ can be performed without actually polling $u$ every time. In particular, in the case of a controller having a centralized view of the network, the backlog $B_{(u,v)}(t)$ can be inferred by the controller from the knowledge of the previously allocated content flows and the link capacity.

Further, we propose Algorithm~\ref{alg:mbp-thresh} to only allocate the elephant flows. As we will see, Algorithm~\ref{alg:mbp-thresh} can achieve most of the gain in response time with only a fraction of the allocation decisions. Algorithm~\ref{alg:mbp-thresh} eases the allocation requirement (only elephant flows need to be represented) but also the amount of information that needs to be kept about content: only elephant flows need to be known.

\begin{algorithm}
\caption{Thresholded MBP algorithm}
\begin{algorithmic}
\REQUIRE $\mathcal P_{s,d}$ for each $(s, d)$ traffic demand pair. 
\STATE $z_{\text{thresh}} \leftarrow$ THRESHOLD,  
\IF{$z_n \ge z_{\text{thresh}}$}
	\STATE {choose $P$ given by the result of Algorithm \ref{alg:mbp}}
\ELSE
	\STATE {choose $P$ randomly among $\mathcal P_{s,d}$ such that MLU is minimized}
\ENDIF
\RETURN path $P$
\end{algorithmic}
\label{alg:mbp-thresh}
\end{algorithm}

\section{Evaluations}
\label{sec:eval}

To evaluate the benefit of Information-Centric Networks on traffic engineering, we compare the Minimum Backlog Policy with the Weighted Random policy for minimized MLU (maximum link utilization), which is a generalization of the Weighted Random policy introduced in section \ref{subsec:moti}. In this section, we first formulate the Weighted Random policy for minimized MLU, so that it can be applied in a general network topology. Subsequently we describe our evaluation set-up and present the results from our event-driven simulations.
\subsection{Formulation of Weighted Random policy for minimized MLU}
To make a valid comparison with the Weighted Random policy, we need to formulate it for a general network topology. Consider traffic demand for each source and destination pair, $(s,d)\in E\times E$. Suppose for each demand of size $\lambda_{s,d}\bar{z}_{s,d}$ from $s$ to $d$, we know the $K_{s,d}$ shortest paths, $P_{s,d}^1, P_{s,d}^2, ..., P_{s,d}^{K_{s,d}}$, a priori and attempt to assign weights, $\pi_{s,d}^1, \pi_{s,d}^2, ..., \pi_{s,d}^{K_{s,d}}$ to them based on which incoming content will be assigned one out of the $K_{s,d}$ paths randomly. We can thus solve the follow linear program to obtain the optimal path weights to achieve the minimized maximum link utilization (dropping the comma within the subscript $\{s,d\}$ of $P, K, \pi, \bar{z}$, and $\lambda$ for clarity):
\begin{equation} \label{eq:rand}
\begin{aligned}
& \text{minimize}
& & t \\
& \text{subject to}
& & \sum_{k=1}^{K_{sd}} \pi_{sd}^k = 1, \forall (s,d). \\
&& & 
\frac{1}{c_{(i,j)}} \sum_{(s,d)}\sum_{k=1}^{K_{sd}} I_{(i,j), P_{sd}^k}\pi_{sd}^k \bar{z}_{sd}\lambda_{sd}  \le t, 
\forall (i,j) \in E. 
\\
&&& t \le 1.
\end{aligned}
\end{equation}
where $I_{(i,j), P_{sd}^k}=1$ when link $(i,j)$ is in path $P_{sd}^k$, and $0$ otherwise. In our simulations, we set $K_{s,d}=3$. This policy works by randomly assign path for a content delivery, based on the weights computed with the linear program above.

\subsection{Simulation set-up and methodology}
\begin{figure}[h]
\begin{center}
\includegraphics[height=5.2cm]{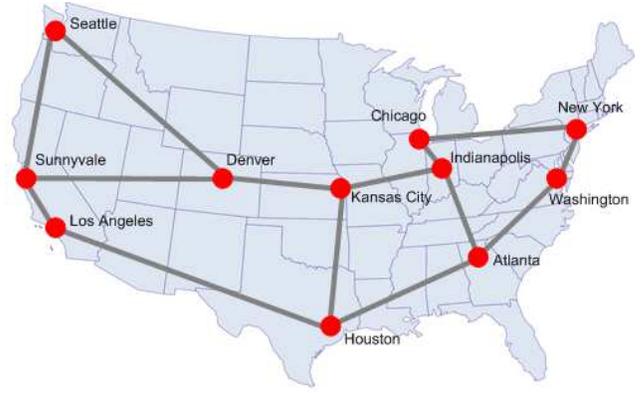}
\end{center}
\caption{Abilene network topology}
\label{fig:topo}
\end{figure}

We implemented the above algorithms with an NS2-based event-driven simulation. We consider Abilene network \cite{abilene} as our topology. The Abilene network is a backbone network established by Internet2 group, and it consists of $11$ nodes (shown in Figure \ref{fig:topo}). We assume the content arrivals for each source destination pair follow a Poisson process. The simulation is performed with two different content size distributions: bimodal and Pareto distribution. We have obtained real life traffic demand matrices from \cite{abilene-tm}. Those matrices were built by measuring the end to end traffic for each source-destination pair in the Abilene network, for a period of $5$ minutes. We have a script to generate content arrival events such that the traffic demands given by aforementioned traffic matrices are satisfied. The capacity for each link is $9920$Mbps except the link between the router in Indianapolis and Atlanta where the capacity is $2480$Mbps. Latency of each link is set to $10$ms. We assign each link in the network an OSPF link weight given by \cite{abilene-tm} and utilize Yen's algorithm \cite{yen} to compute $3$ shortest paths for each source-destination traffic demand based on the link weights. As mentioned in section \ref{sec:design-space}, we are mostly interested to see a minimum backlog driven flow allocation policy, MBP, could reduce the mean response time of content deliveries. We measure the response time of a piece of transferred content, by calculate the difference between the arrival of the acknowledgement for the last packet in the content, and the arrival of the content.

\subsection{Simulation results}
\subsubsection{Mean response time improvement}\label{sec:mrt} We chose a traffic matrix for the real network traffic in May 15th, 2004. Solving equations (\ref{eq:rand}) for this traffic matrix yields a resulting MLU of $0.603$. We are interested to see the gains in reducing mean response time by applying Minimum Backlog Policy against the optimal weighted random policy, with respect to MLUs from $0.6$ towards $1$. We thus simulate content deliveries with the two different path selection policies respectively, with the traffic matrix multiplied by a scalar, ranging from $1.0$ to $1.6$. The resulting MLU given by equations (\ref{eq:rand}) are from $0.603$ to $0.964$. Figures 3 and 4 correspond to simulations with two different content size distributions: Pareto and bimodal, respectively. To characterize the response time improvement, we define the response time reduction gain as the ratio of the response time difference against the response time of weighted random policy. Both simulations show that, MBP reduces at least $28\%$ the mean response time, compared with weighted random policy (or conversely, min-MLU is at least 39\% slower). This response time reduction gain increases when the network is more heavily loaded, i.e., when the optimal MLU grows from $0.60$ to $0.96$.

\begin{figure}[t] \label{fig:pareto}
\begin{center}
	\subfigure[Mean response time comparison] {
		\label{fig:mrt-cmp-p}
		 \includegraphics[width=4cm]{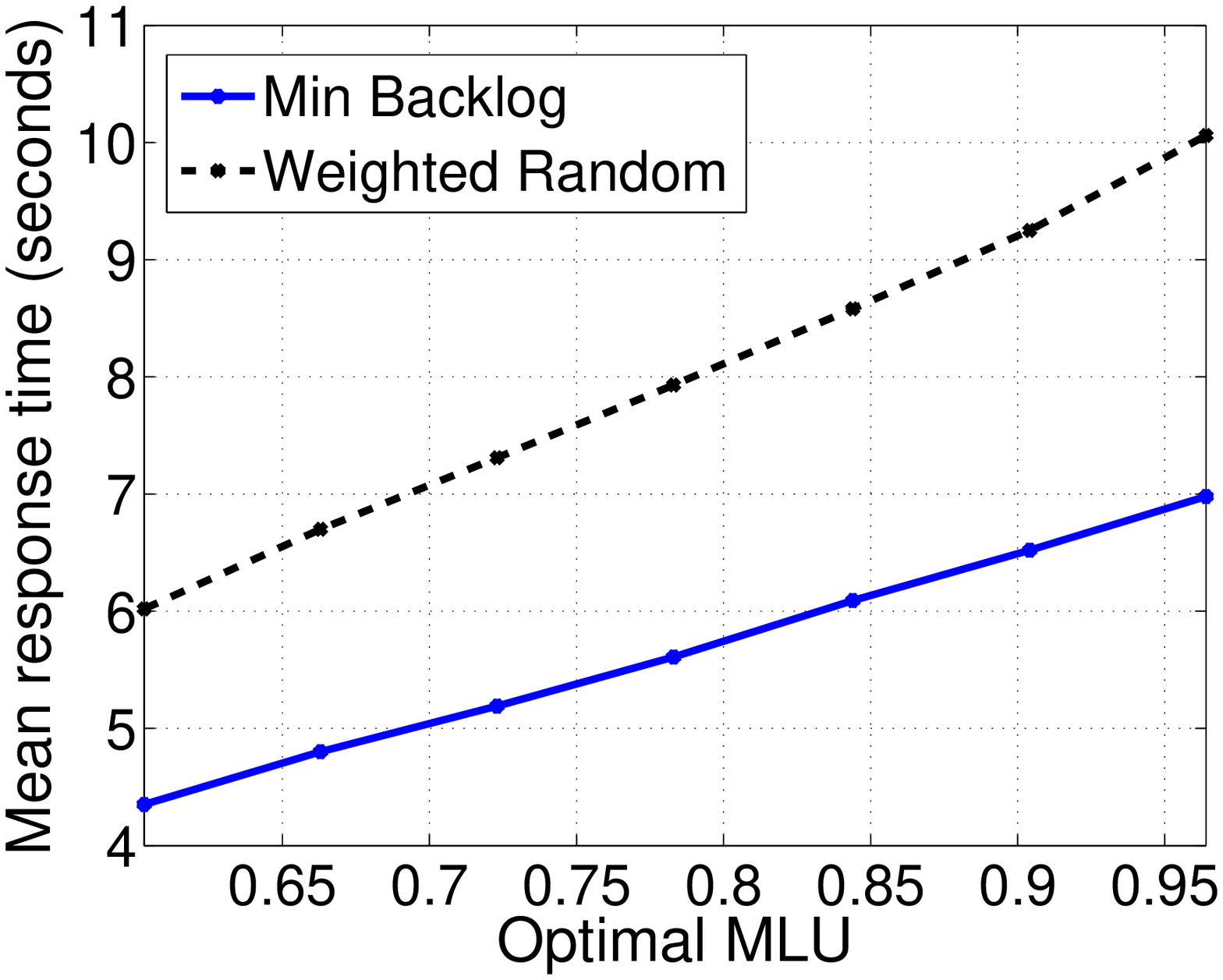}
	}
	\subfigure[Response time reduction gain] {
		\label{fig:mrt-gain-p}
		 \includegraphics[width=4cm]{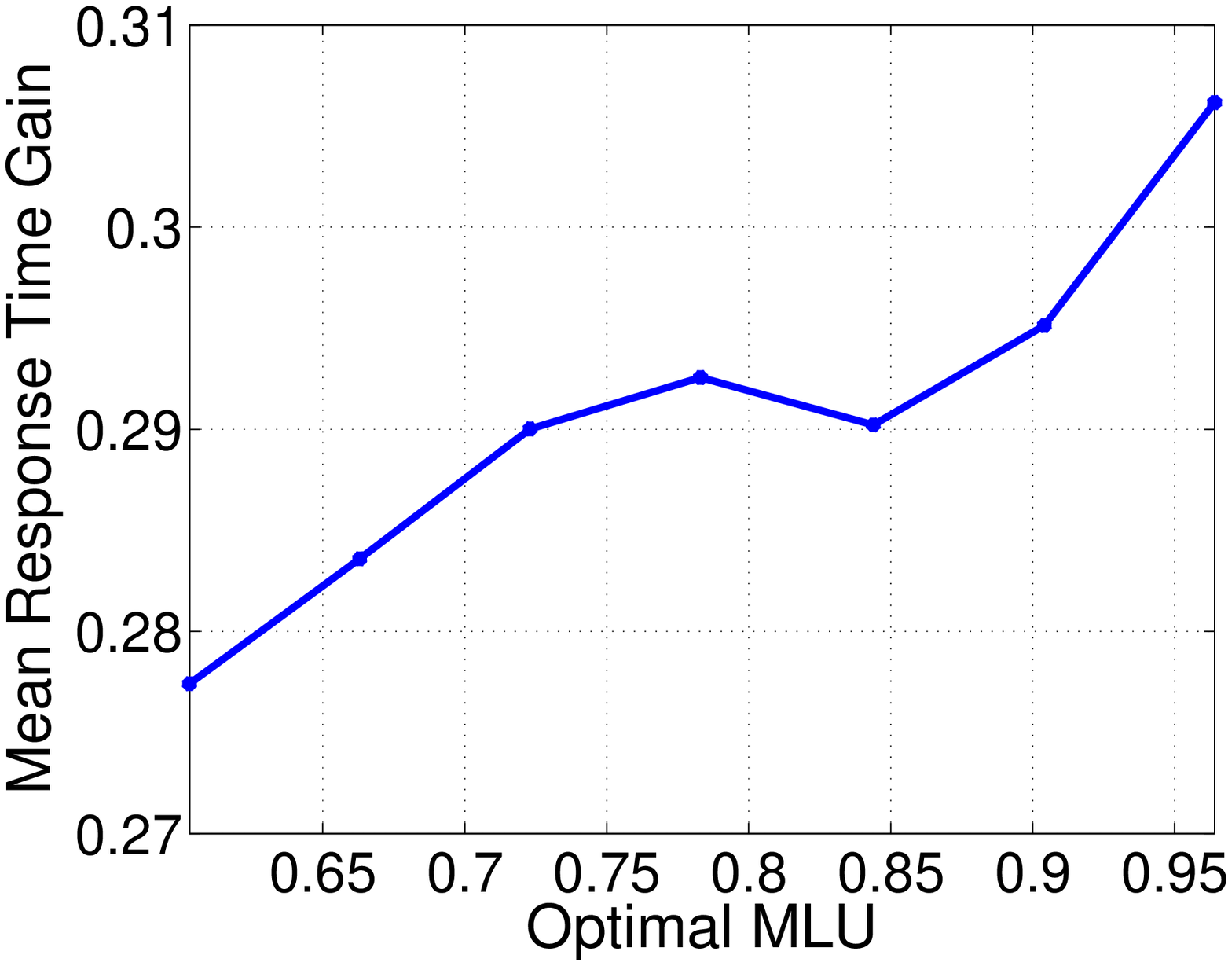}
	}
\caption{Mean response time comparison for the two policies, with traffic matrices whose optimal MLU range from 0.6 towards 1.0, and content sizes following \textit{Pareto} distribution.}
\end{center}
\end{figure}

\begin{figure}[t]
\label{fig:bimodal}
\begin{center}
	\subfigure[Mean response time comparison] {
		\label{fig:mrt-cmp-b}
		 \includegraphics[width=4cm]{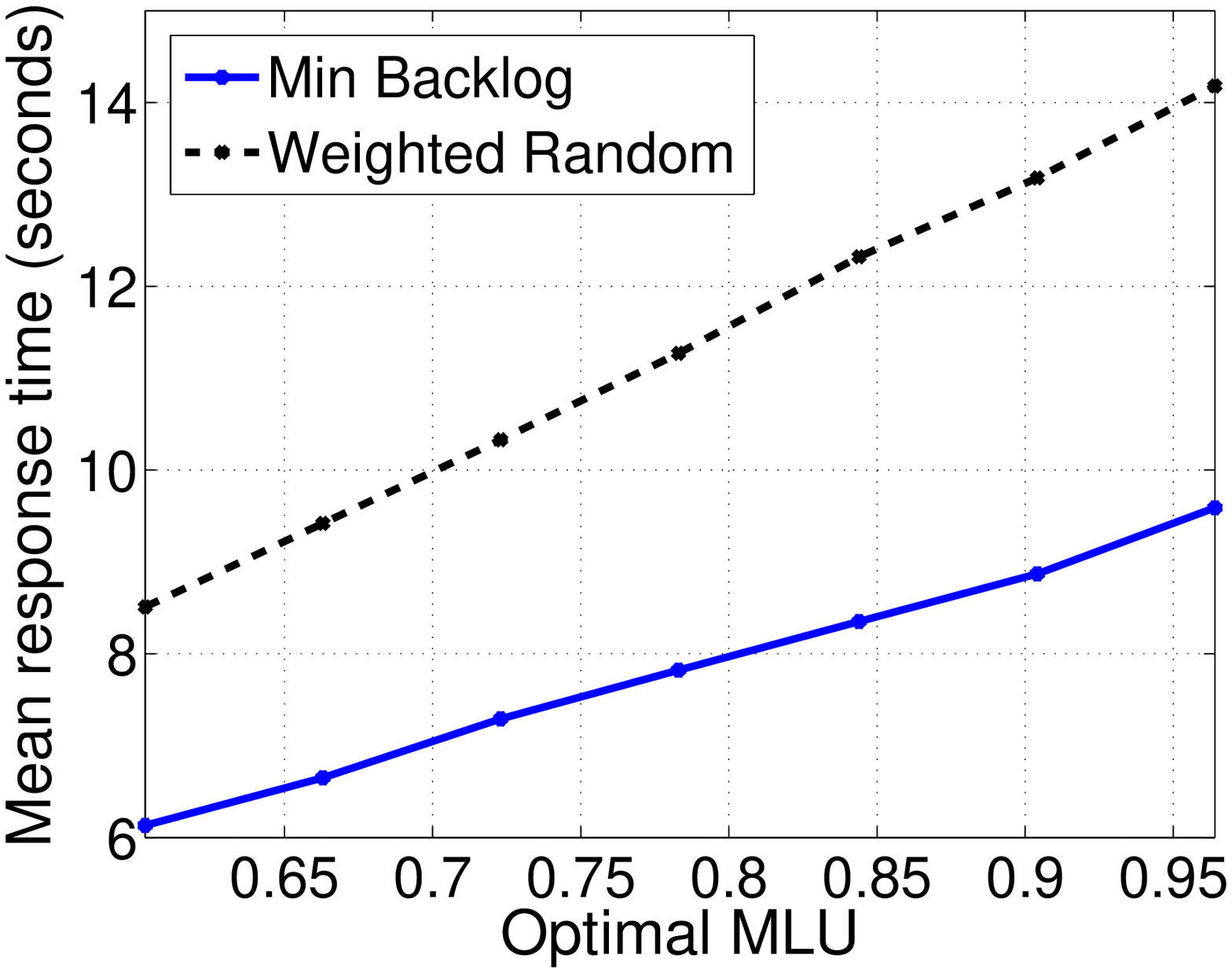}
	}
	\subfigure[Response time reduction gain] {
		\label{fig:mrt-gain-b}
		 \includegraphics[width=4cm]{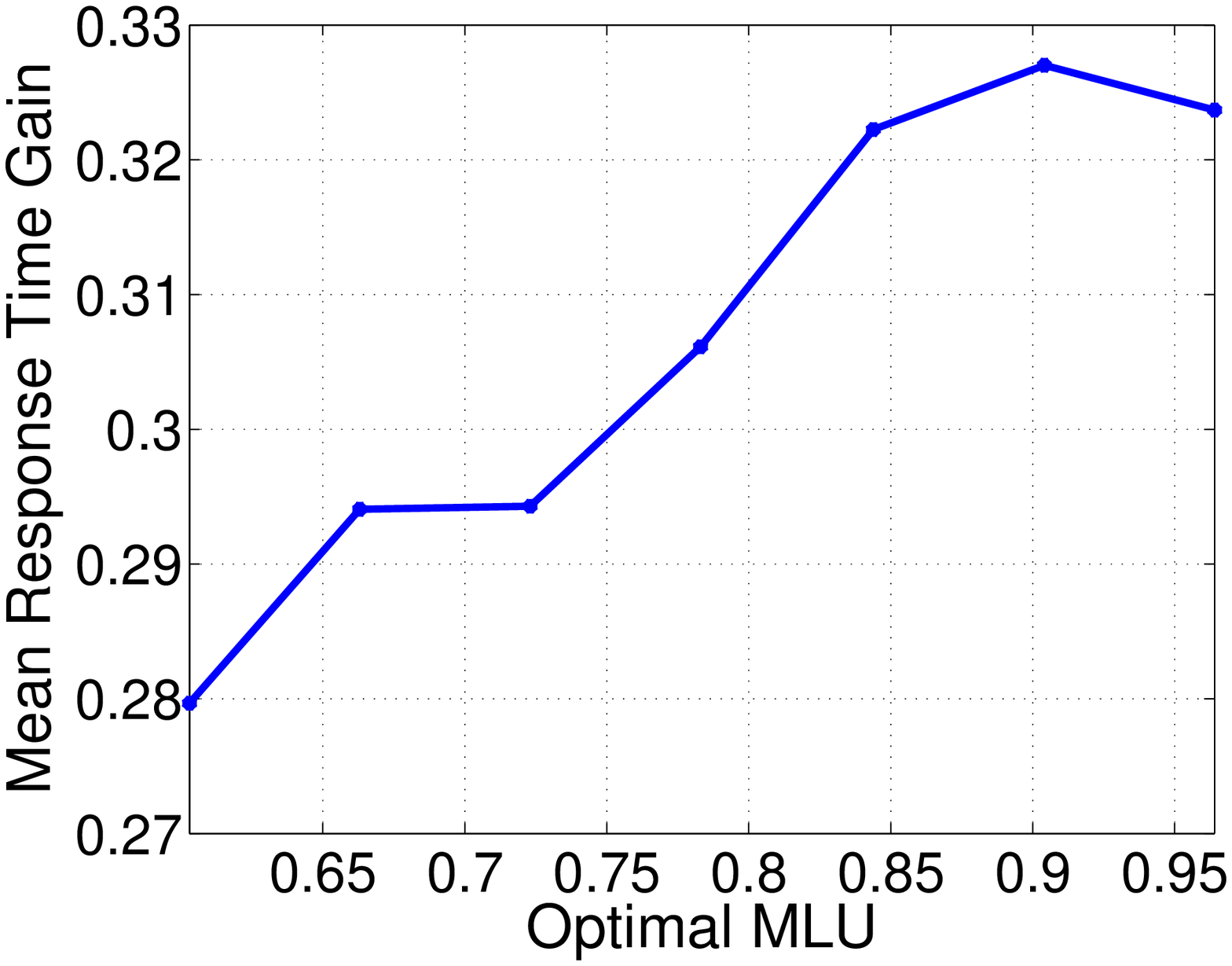}
	}
\caption{Mean response time comparison for the two policies, with traffic matrices whose optimal MLU range from 0.6 towards 1.0, and content sizes following \textit{Bimodal} distribution.}
\end{center}
\end{figure}

\subsubsection{On thresholding the content size for MBP allocations}

The results shown above correspond to the design decision that each incoming flow, no matter how small its size, would require an MBP based allocation. This is certainly optimal, in terms of reducing the response time. However, allocating based on MBP for every flow is practically undesirable: flows of tiny sizes themselves are unlikely to congest the network, and yet path allocations would incur network operational overheads, such as computation, and message passing. The key is to identify a proper threshold of flow size,  performing MBP based allocation only for flows of sizes above the threshold and randomly assign based on computed optimal weights for the rest. 

\begin{figure}[t]
\label{fig:thresh}
\begin{center}

	\subfigure[Response time reduction gain] {
		\label{fig:thresh-mrt}
		 \includegraphics[width=6cm]{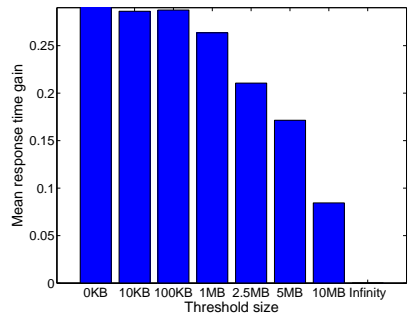}
	}
	\subfigure[Allocation frequencies] {
		\label{fig:thresh-freq}
		 \includegraphics[width=6cm]{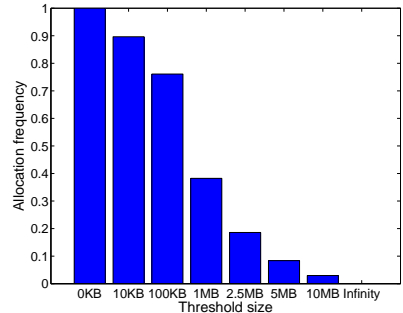}
	}
\caption{Performance of thresholded MBP, with the traffic matrix whose optimal MLU is $0.783$. The considered thresholds range from 10KB to 10MB. A threshold of 0 and infinity reduce to Min Backlog and Weighted Random, respectively. }
\end{center}
\end{figure}

To explore this direction, we take the traffic matrix introduced in section \ref{sec:mrt} scaled by $1.3$, and carry out the simulation with thresholded MBP allocation of Algorithm~\ref{alg:mbp-thresh}. The tested thresholds range from $10$KB to $10$MB. In essence, the MBP and weighted random policy are equivalent to setting the thresholds to $0$KB and infinity, respectively. We thus plot the response time gains and allocation frequencies for thresholds from $0$ to infinity. Note that the allocation frequency is defined as the ratio of number of MBP allocations against the total number of the flows. Figures \ref{fig:thresh-mrt} and \ref{fig:thresh-freq} show that the gain decreases as the threshold value is enlarged, along with the reduced allocation frequency. Additionally, it can be seen from the figure that a threshold of $2.5$MB only requires $20\%$ allocations, but can still maintain at least a response time reduction gain of $20\%$.
\section{Conclusion}
\label{sec:conclusion}

We have presented a traffic engineering policy for Information-Centric Networks which leverages the fact that Information-Centric Network can be made aware of properties of the content, such as their size. We have seen that nodes (or a centralized controller) can keep an aggregate value of the backlog on their links, and the network can use this information for performing dynamic traffic engineering over new content arrivals into the network. We hope that it demonstrates that a significant benefit of ICNs lie in their tighter granularity in describing the network content, and that content knowledge can improve network performance and perform better resource allocation. 

We have presented the MBP policy which allocates traffic to paths with the least amount of backlog as well as a variant which only allocates elephant flows above a threshold. We see that both policies significantly improves the network's response time for a wide range of network utilization and for different traffic size distributions (namely, bi-modal and Paretto). Using both a simple intuition-building topology and the Abilene network topology, we have thoroughly evaluated the proposed policy and shown reduced delay in all cases, and up to 33\% improvement over min-MLU (that is, min-MLU is 50\% slower) in some evaluation scenarios. 

\bibliographystyle{IEEEtran}
\bibliography{library}

\end{document}